\documentclass[11pt]{article}
\usepackage{amsfonts,amsmath,amssymb,graphics,epsfig}

\begin{document}

\date{\empty}

\title{\textbf{Cosmological magnetic field survival}}

\author{John D. Barrow$^1$ and Christos G. Tsagas$^2$\\ {\small $^1$DAMTP, Centre for Mathematical Sciences}\\ {\small University of Cambridge, Wilberforce Road, Cambridge CB3 0WA, UK}\\ {\small $^2$Section of Astrophysics, Astronomy and Mechanics, Department of Physics}\\ {\small Aristotle University of Thessaloniki, Thessaloniki 54124, Greece}}
\maketitle

\begin{abstract}
It is widely believed that primordial magnetic fields are dramatically diluted by the expansion of the universe. As a result, cosmological magnetic fields with residual strengths of astrophysical relevance are generally sought by going outside standard cosmology, or by extending conventional electromagnetic theory. Nevertheless, the survival of strong $B$-fields of primordial origin is possible in spatially open Friedmann universes without changing conventional electromagnetism. The reason is the hyperbolic geometry of these spacetimes, which slows down the adiabatic magnetic decay-rate and leads to their superadiabatic amplification on large scales. So far, the effect has been found to operate on Friedmannian backgrounds containing either radiation or a slow-rolling scalar field. We show here that the superadiabatic amplification of large-scale magnetic fields, generated by quantum fluctuations during inflation, is essentially independent of the type of matter that fills the universe and appears to be a generic feature of open Friedmann spacetimes. We estimate the late-time strength of any residual field in a marginally open universe and show that it can easily meet the requirements for the dynamo generation of the magnetic fields observed in galaxies today.\\\\ PACS numbers: 98.80.-k, 98.62.En, 98.65.Dx
\end{abstract}

\section{Introduction}\label{sI}
Magnetic fields are common in many astrophysical environments. From the Earth, the Sun and the Milky Way, up to distant galaxies, galaxy clusters and high-redshift protogalactic structures, the presence of magnetic fields has been repeatedly verified~\cite{U}. Galaxies, galactic clusters and protogalactic clouds, in particular, are known to host coherent magnetic fields between $\sim10^{-7}$ and $\sim10^{-6}$~Gauss. Yet, despite their ubiquity, the origin of these large-scale magnetic fields is still largely unknown~\cite{GR}. It is generally believed that the galactic $B$-fields have been amplified and sustained by some kind of dynamo action~\cite{KA}. Nevertheless, even if we bypass questions regarding its efficiency, the galactic dynamo requires an initial seed-field in order to operate. It is harder still to explain the magnetic fields that have been observed in galaxy clusters and in remote protogalactic clouds~\cite{Ketal}, as well as recent observations reporting $B$-fields close to $10^{-15}$~G in empty intergalactic space~\cite{TGFBGC}.

Typically, the possible mechanisms for producing the large-scale magnetic fields seen in the universe today are classified into those operating after recombination and those that seek an early cosmological origin~\cite{GR}. The idea of primordial magnetism is attractive because it could readily explain the widespread presence of magnetic fields in the cosmos, especially those seen in high-redshift protogalaxies and in empty intergalactic voids. Nevertheless, early-time magnetogenesis has its problems. The main theoretical obstacle appears to be the very weak residual strength of cosmological $B$-fields. As a result, magnetic fields of primordial origin in Friedmann-Robertson-Walker (FRW) universes have been largely treated as astrophysically irrelevant.

The root of the problem can be traced to the widespread belief that, on FRW backgrounds, magnetic fields always decay adiabatically, with $B\propto a^{-2}$ (where $a=a(t)$ is the cosmological scale-factor). This means magnetic strengths below $10^{-50}$~G today. To the best of our knowledge, such fields can never seed the galactic dynamo, or affect the dynamical evolution of our universe. However, the literature makes no clear distinction between flat and curved Friedmann models. This has led to the belief that the adiabatic magnetic decay-law holds in all three FRW universes, irrespective of their spatial geometry. As a result, solutions to the aforementioned magnetic-strength problem are generally sought outside what one might call standard cosmology or conventional electromagnetic theory. There are many different ways of doing that, which explains the large number and the variety of the scenarios proposed in the literature  (see~\cite{TW} for a representative, but incomplete, list).

The adiabatic magnetic decay in FRW cosmologies results from two theoretical facts: the conformal invariance of Maxwellian electromagnetism and the conformal flatness of the Friedmannian spacetimes. The two are thought to guarantee the Minkowski-like, $B\propto a^{-2}$ magnetic decay law in all FRW models at all times (e.g.~see~\cite{GR} for reviews and also~\cite{sub1} for a more technical discussion). This is true, to linear order, but (strictly speaking) only when the Friedmann background has Euclidean spatial sections. Although all three FRW spacetimes are conformally flat, only those with Euclidean 3-geometry are \emph{globally} conformal to the Minkowski spacetime. For the non-flat Friedmann universes, the conformal mappings are \emph{local}~\cite{S}. In other words, the conformal factor of the spatially curved FRW spacetimes no longer coincides with the cosmological scale factor, but has an additional spatial dependence. This means that the simple $B\propto a^{-2}$ evolution law, characteristic of the flat Friedmann universes, does not generally hold in FRW models with non-Euclidean space sections. In fact, when the spatial geometry is hyperbolic, the 3-curvature can slow down the adiabatic decay of the field and lead to its superadiabatic amplification.\footnote{``Superadiabatic amplification'' does not necessarily imply an actual increase in the strength of the $B$-field. Instead, the term has been used in the literature to describe magnetic decay-rates slower than the standard (adiabatic -- $B\propto a^{-2}$) rate: there is no photon production during this process.} Not surprisingly, the effect occurs on large scales, where the curvature of the space plays a prominent role. More specifically, $B$-fields spanning lengths close to the curvature radius of an open Friedmann universe were found to decay as $a^{-1}$ throughout the radiation era and also during a phase of slow-roll inflation~\cite{TK}. This can lead to residual $B$-fields substantially stronger than previously anticipated and strong enough to seed a galactic dynamo mechanism.

Here, we revisit and extend the scenario proposed in~\cite{TK} and show that the superadiabatic amplification of large-scale magnetic fields is essentially independent of the type of matter that fills the universe. To a large extent, this makes the aforementioned amplification effect a generic feature of the open FRW spacetimes. We also re-estimate the final strength of the superadiabatically amplified $B$-field and find that, in almost all the cases of interest, it can readily seed the galactic dynamo. More specifically, for a substantially broad spectrum of initial conditions, the typical magnitude of the residual $B$-field varies between $10^{-20}$ and $10^{-10}$~Gauss. Overall, and contrary to general belief, it appears that Friedmann cosmologies can naturally sustain primordial magnetic fields of astrophysical interest, without introducing non-standard variants of classical electromagnetism, if they have hyperbolic spatial geometries.

\section{Linear magnetic evolution on FRW 
backgrounds}\label{sLMEFRWBs}
In order to understand how spatial curvature can influence and
superadiabatically amplify large-scale magnetic fields in open Friedmann models, we first need to investigate the linearised magnetic evolution on general FRW backgrounds.

\subsection{Maxwell's equations}\label{ssMEs} 
The starting point is Maxwell's equations, which monitor the electromagnetic field. Relative to a fundamental observer moving with 4-velocity $u_{a}$ (normalised so that $u_{a}u^{a}=-1$), they split into a pair of propagation equations.\footnote{Given the spacetime metric ($g_{ab}$) and a family of timelike worldlines tangent to the 4-velocity field $u_{a}$, we define the tensor $h_{ab}=g_{ab}+u_{a}u_{b}$. The latter projects orthogonal to $u_{a}$ and into the observers' instantaneous 3-dimensional rest-space. Then, on using $u_{a}$ and $h_{ab}$, every variable, every operator and every equation can be decomposed into their irreducible timelike and spacelike components~\cite{TB,TCM}.} One for the electric ($E_{a}$) field
\begin{equation}
\dot{E}_{\langle a\rangle}= -{\frac{2}{3}}\,\Theta E_{a}+ \left(\sigma_{ab}+\omega_{ab}\right)E^{b}+ \varepsilon_{abc}A^{b}B^{c}+\mathrm{curl}B_{a}- \mathcal{J}_{a}\,, \label{Edot}
\end{equation}
and another one for its magnetic ($B_{a}$) counterpart
\begin{equation}
\dot{B}_{\langle a\rangle}= -{\frac{2}{3}}\,\Theta B_{a}+ \left(\sigma_{ab}+\omega_{ab}\right)B^{b}- \varepsilon_{abc}A^{b}E^{c}- \mathrm{curl}E_{a}\,,  \label{Bdot}
\end{equation}
with the overdots indicating proper-time derivatives along the worldlines of the fundamental observers~\cite{TB,TCM}.\footnote{Angled brackets denote the symmetric and trace-free component of tensors and the orthogonally projected part of vectors (e.g.~$\dot{E}_{\langle a\rangle}= h_{a}{}^{b}\dot{E}_{b}$). Square brackets, on the other hand, indicate antisymmetric tensors. Also, $\varepsilon_{abc}$ represents the Levi-Civita tensor of the 3-dimensional hypersurface orthogonal to $u_{a}$ and $\mathrm{curl}v_{a}= \varepsilon_{abc}\mathrm{D}^{b}v^{c}$ defines the ``curl'' of a given spacelike vector $v_{a}$. Finally, $\mathrm{D}_{a}= h_{a}{}^{b}\nabla _{b}$ is the covariant derivative operating in the observers 3-dimensional rest-space, with $\nabla_{a}$ being the covariant derivative of the whole spacetime~\cite{TB,TCM}.} The kinematics of these worldlines trigger relative-motion effects, which are encoded in the first three terms on the right-hand side of the above. The scalar $\Theta=\nabla^{a}u_{a}=\mathrm{D}^{a}u_{a}$ describes changes in the observers' average separation (i.e.~expansion or contraction), while the tensors $\sigma_{ab}=\mathrm{D}_{\langle a}u_{b\rangle}$ and $\omega_{ab}=\mathrm{D}_{[a}u_{b]}$ represent kinematic changes due to shear and vorticity distortions, respectively. Also, the vector $A_{a}=\dot{u}_{a}=u^{b}\nabla_{b}u_{a}$ is the 4-acceleration and reflects the presence of non-gravitational forces. Finally, $\mathcal{J}_{a}$ is the spatial electric current, which is related to the electric field via Ohm's law~\cite{G}:
\begin{equation}
\mathcal{J}_{a}= \varsigma E_{a}\,.  \label{Ohm}
\end{equation}
The scalar $\varsigma$ depends on the electric properties of the matter and varies between the two opposite limits of very high (i.e.~$\varsigma\rightarrow\infty$) and very low (i.e.~$\varsigma\rightarrow0$) electrical conductivity. In the former case one has the ideal magnetohydrodynamic (MHD) approximation, where the electric fields vanish despite the presence of nonzero currents. In poorly conductive environments, on the other hand, the situation is reversed. To a certain extent, the form of Eqs.~(\ref{Edot}) and (\ref{Bdot}) and thus the evolution of cosmological $B$-fields, is decided by the electric properties of the cosmic medium.

\subsection{The case of high conductivity}\label{ssCHC} 
It is generally believed that the universe has been a very good electrical conductor throughout its standard Hot-Big-Bang evolution, namely through the radiation and the dust eras. During these periods, and on scales well within the horizon, the currents have essentially obliterated any pre-existing large-scale electric fields and kept the magnetic fields frozen into the highly conductive plasma. Consequently, assuming an FRW background of high electrical conductivity (i.e.~setting $E_{a}=0$ and $A_{a}=0=\sigma_{ab}= \omega_{ab}
$), expression (\ref{Bdot}) linearises to
\begin{equation}
\dot{B}_{a}= -2HB_{a}\,,  \label{MHDB}
\end{equation}
where $H=\Theta/3=\dot{a}/a$ is the (zero order) Hubble parameter. This ensures that $B\propto a^{-2}$ at all times. In other words, as long as the ideal-MHD approximation holds, the magnetic flux is conserved and the $B$-field decays adiabatically, irrespective of the equation of state of the matter and of the spatial geometry of the FRW host.

\subsection{The case of low conductivity}\label{ssCLC} 
In contrast to the high conductivity of the post-inflationary era, the universe was a very poor electrical conductor during inflation. The situation is then reversed: we can ignore the electric currents but now need to take the electric fields into account. This is also the case after inflation on scales far beyond the Hubble radius, where the conductivity remains very low. The reason is causality, which confines the highly conductive currents of the post-inflationary era well inside the horizon.\footnote{Causality is limited by the observer's particle horizon, which in many FRW models essentially coincides with the Hubble radius. However, in Friedmann universes with hyperbolic spatial sections, the particle horizon can exceed the Hubble scale (e.g.~see~\cite{W}). This means that wavelengths larger than the curvature radius of an open FRW cosmology can be causally connected, despite the fact that they always lie outside the Hubble length. Here, to keep things simple, we will still treat the Hubble scale as our causal horizon as
well.} In poorly conducting environments, the magnetic evolution is no longer determined by Eq.~(\ref{Bdot}) alone, but by the coupled set of (\ref{Edot}) and (\ref{Bdot}). These combine to a wave equation for the magnetic (and also the electric) field, which when linearised on a FRW background reads~\cite{T1}
\begin{equation}
\ddot{B}_{a}- \mathrm{D}^{2}B_{a}= -5H\dot{B}_{a}- 2\left( \dot{H}+3H^{2}+a^{-2}K\right)B_{a}\,,  \label{Bddot}
\end{equation}
with $\mathrm{D}^{2}=\mathrm{D}^{a}\mathrm{D}_{a}$ representing the
3-dimensional covariant Laplacian operator and $K=0,\pm 1$ the background 3-curvature index. For our purposes, the key term is the magneto-geometrical on the right-hand end of the above expression. This term, which results from the Ricci identities, reflects the fact that we are dealing with a source of vector nature (the magnetic field) within a geometrical theory of gravity. On using conformal (i.e.~$\eta $ with $\dot{\eta}=1/a$) instead of proper
time, Eq.~(\ref{Bddot}) simplifies to
\begin{equation}
\mathcal{B}_{a}^{\prime\prime}- a^{2}\mathrm{D}^{2}\mathcal{B}_{a}= -2K\mathcal{B}_{a}\,,  \label{cBddot1}
\end{equation}
where $\mathcal{B}_{a}=a^{2}B_{a}$. Finally, we introduce the harmonic splitting $\mathcal{B}_{a}= \Sigma_{n}\mathcal{B}_{(n)}\mathcal{Q}_{a}^{(n)}$, where $\mathcal{B}_{(n)}$ is the $n$-th magnetic mode (with $\mathrm{D}_{a}\mathcal{B}_{(n)}=0$) and $n$ is the associated comoving eigenvalue.\footnote{The eigenvalues have a continuous spectrum, with $n^{2}\geq0$, in spacetimes with Euclidean or hyperbolic spatial sections and a discrete one, with $n^{2}\geq2$, when the 3-dimensional hypersurfaces are positively curved.} Also, $\mathcal{Q}_{a}^{(n)}$ are pure-vector harmonics that satisfy the conditions $\mathcal{Q}_{a}^{\prime\,(n)}= 0=\mathrm{D}^{a}\mathcal{Q}_{a}^{(n)}$ and the vector version of the Laplace-Beltrami equation, that is
\begin{equation}
\mathrm{D}^{2}\mathcal{Q}_{a}^{(n)}= -\left({n\over a}\right)^{2}\mathcal{Q}_{a}^{(n)}\,.  \label{LB}
\end{equation}
Applying the above decomposition to Eq.~(\ref{cBddot1}), the harmonics decouple and the wave formula of the $n$-th magnetic mode assumes the form
\begin{equation}
\mathcal{B}_{(n)}^{\prime\prime}+ n^{2}\mathcal{B}_{(n)}= -2K\mathcal{B}_{(n)}\,.  \label{cBddot2}
\end{equation}
The latter describes the linear evolution of large-scale magnetic
fields on a general FRW background, provided the matter fields involved are poor electrical conductors. According to (\ref{cBddot2}), the magnetic evolution depends on the spatial geometry of the unperturbed universe. This is especially important for $B$-fields coherent on scales close to and beyond the curvature radius of the background model, which corresponds to the $n^{2}=1$ threshold~\cite{LS}.

\section{Adiabatic magnetic decay in flat FRW 
universes}\label{sAMDFFRWUs}
In order to operate successfully, a galactic dynamo requires seed-fields that are coherent on comoving lengths no less that 10~Kpc. This means that magnetic fields generated between inflation and (roughly) recombination (when $z\simeq1100)$ are typically too small and will not help the amplification process. This is known as the ``scale problem''.

\subsection{ Magnetic fields produced by inflation}\label{ssIPMFs}
The root of the scale problem is causality, because it confines the coherence length of the $B$-field inside that of the horizon at the time of magnetogenesis. For instance, magnetic seeds produced during the electroweak phase-transition (with $z\sim 10^{14}$) have current lengths of the order of 1~AU only. One can increase the coherence size of the original field by transferring magnetic energy from smaller to larger scales, through an ``inverse cascade'' mechanism~\cite{BEO}. The latter, however, requires rather significant amounts of magnetic helicity in order to operate efficiently.

Inflation has long been considered as potentially the best mechanism for solving the aforementioned scale-problem, since it naturally creates large-scale correlations from microphysical processes that operate within the Hubble radius. The inflationary picture also provides the dynamical means of producing long-wavelength electromagnetic perturbations, by stretching  small-scale quantum mechanical fluctuations in the Maxwell field to super-Hubble scales. Once outside the Hubble horizon, these quantum-mechanically excited electromagnetic modes are expected to freeze-in as classical, static electromagnetic waves. The latter will lead to current-supported magnetic fields when the modes in question re-enter the Hubble radius during the subsequent radiation and dust eras. At that time, the highly conductive currents will also eliminate the electric fields, leaving the universe permeated by a large-scale magnetic field of cosmological origin.

Nevertheless, although $B$-fields that have survived a phase of early inflationary expansion can solve the scale-problem, they are generally too weak to seed the galactic dynamo. As we will now see, the reason is the adiabatic magnetic decay law.

\subsection{Typical residual strength of inflationary magnetic
fields}\label{ssTRSIMFs}
When the unperturbed FRW model has flat spatial sections, the right-hand side of (\ref{cBddot2}) vanishes and the latter assumes a simple Minkowskian form. The resulting differential equation accepts an oscillatory solution, which (written in terms of the actual magnetic field $B=\mathcal{B}/a^2$) reads
\begin{equation}
B_{(n)}= \left[ \mathcal{C}_{1}\sin(n\eta)+ \mathcal{C}_{2}\cos\left(n\eta\right)\right] \left({\frac{a_{0}}{a}}\right)^{2}\,,  \label{fB}
\end{equation}
with the integration constants determined by the initial conditions. Hence, as long as the background spatial geometry is Euclidean, the overall adiabatic magnetic decay-rate is preserved. Combining this result with the one obtained previously, for the case of high electrical conductivity, we see that magnetic fields decay adiabatically throughout the entire evolution of the universe (i.e.~during inflation, radiation and dust) on all scales. The immediate consequence is that, on spatially-flat FRW backgrounds, primordial $B$-fields are catastrophically diluted by the universal expansion and of no astrophysical interest today.

To calculate the extent of the magnetic dilution, recall that at first horizon crossing, the relative strength of the $n$-th magnetic mode is $(\rho_{B}/\rho)_{HC}\simeq f_Q(M/M_{Pl})^{4}$, where $\rho_{B}=B_{(n)}^{2}$ is the magnetic energy density, $\rho$ is that of the background matter component and $M_{Pl}\simeq10^{19}$GeV is the Planck mass. Note that we have also allowed for the dimensionless parameter $f_Q$, which in a sense measures the efficiency of the quantum mechanical excitation of the electromagnetic modes. In practice, $f_Q$ has always been assumed to equal unity, though generally one expects that $f_Q\in[0,1]$. Now, during the de Sitter phase we have $\rho\simeq M^{4}\simeq$~constant, with $M$ representing the energy scale of the adopted inflationary scenario. Then, the adiabatic magnetic decay-law (see solution (\ref{fB}) above) implies that $B_{(n)}^{2}=(B_{(n)}^{2})_{HC}\,\mathrm{e}^{-4N}$ by the time inflation is over, where $N$ is the number of e-folds between horizon-crossing and the end of the inflationary expansion. Overall, recalling that $(\rho_B/\rho)_{RH}\simeq (\rho_B/\rho)_{INF}(T_{RH}/M)^{4/3}$ is the relative change in the magnetic energy density between the end of inflation proper and that of reheating, we find that at the onset of the radiation era
\begin{equation}
r_{RH}= \left({\frac{\rho _{B}}{\rho_{\gamma }}}\right)_{RH}\simeq 10^{-104} f_Q\left(\frac{\rm 1Mpc}{\lambda_{B}}\right)^{4}\,,  \label{r1}
\end{equation}
where $\rho_{\gamma}\simeq\rho_{RH}\simeq T^4_{RH}$ is the energy density of the relativistic species and $\lambda_{B}$ is the current scale of the magnetic field~\cite{TW}. The above ratio is independent of the energy scale of the inflationary model, as well as the associated reheat temperature, and remains unchanged throughout the subsequent evolution of the universe (since $\rho_{B}$, $\rho_{\gamma}\propto a^{-4}$ after inflation). In order to operate successfully, the galactic dynamo requires the $B$-seeds to be coherent over a minimum (comoving) scale of approximately 10~Kpc. On these lengths, expression (\ref{r1}) translates to a maximum value of $r_0=(\rho_{B}/\rho_{\gamma})_0\sim10^{-96}$ today. Finally, recalling that $(\rho_{\gamma})_0\sim10^{-51}~\mathrm{GeV}^{4}$, we obtain $B_0\sim 10^{-53}$~G for the corresponding maximum magnetic field strength. Such $B$-fields can never seed the galactic dynamo and are astrophysically unimportant.

\section{Superadiabatic magnetic amplification in open FRW 
universes}\label{sS-AMAOFRWUs}
Over the last twenty years there have been many attempts to find mechanisms that can reduce the adiabatic magnetic decay-rate and thus lead to substantially stronger large-scale $B$-fields. As stated earlier, most of the proposed scenarios achieve their goal by introducing new physics.

\subsection{The physics of the amplification mechanism}\label{ssPAM} 
The main reason for breaking away from conventional physics appears to be the belief that the adiabatic magnetic decay is guaranteed in all FRW cosmologies at all times. In Friedmann models with non-Euclidean spatial sections, however, the magnetic wave equation has an additional curvature-related term. The latter could in principle slow down the adiabatic magnetic decay and produce considerably stronger residual $B$-fields, without abandoning standard electromagnetism or general relativity. This is what happens on FRW backgrounds with hyperbolic spatial geometry,
where Eq.~(\ref{cBddot2}) reads
\begin{equation}
\mathcal{B}_{(n)}^{\prime\prime}+ \left(n^{2}-2\right)\mathcal{B}_{(n)}= 0\,,  \label{cBddot3}
\end{equation}
with $n^{2}\geq0$. This reveals a change in the nature of the magnetic evolution at the $n^{2}=2$ threshold. In particular, recalling that $n^{2}=1$ corresponds to the curvature radius of the universe, we see that the effects of the hyperbolic spatial geometry are negligible on small enough lengths (i.e.~for $n^{2}>2$). There, solution (\ref{fB}) still holds and the adiabatic decay of the magnetic field persists. On large wavelengths (i.e.~with $n^{2}<2$), however, this is no longer the case. The 3-curvature effects are not negligible there and Eq.~(\ref{cBddot3}) has a solution of the form
\begin{equation}
\mathcal{B}_{(k)}= \mathcal{C}_{3}\cosh\left(|k|\eta\right)+ \mathcal{C}_{4}\sinh\left(|k|\eta\right)\,.  \label{ocB}
\end{equation}
Observe that for notational convenience we have introduced the $k$-parameter, defined so that $k^{2}=2-n^{2}$ and $0<k^{2}<2$~\cite{TK}. Therefore, $k^{2}=1$ corresponds to the 3-curvature scale, which is where the effects of the non-Euclidean spatial geometry begin to dominate the FRW dynamics~\cite{LS}. Wavelengths with $0<k^{2}<1$ correspond to the largest subcurvature modes (with $1<n^{2}<2$), while the values $1<k^{2}<2$ are associated with the supercurvature lengths (where $0<n^{2}<1$).\footnote{Following~\cite{LK}, the eigenvalue ($n$) and the wavenumber ($\nu $) of a given harmonic mode are related by $n^{2}=\nu ^{2}-2K$, where $K=0,\pm 1$ is the curvature parameter of the spatial sections. With this definition, magnetic modes having eigenvalues in the $0<n^{2}<2$ range correspond to wave functions with imaginary wavenumbers. In quantum-cosmology nomenclature such modes are usually termed supercurvature~\cite{ST}. Nevertheless, in our case, half of the corresponding scales (those with $1<n^{2}<2$) are smaller than the curvature radius of the universe~\cite{LS}. We will therefore use the label ``supercurvature'' only for modes having $0<n^{2}<1$.} In terms of the actual magnetic field ($B=\mathcal{B}/a^{2})$, expression (\ref{ocB}) reads
\begin{equation}
B_{(k)}= \left(\mathcal{C}_{5}e^{|k|\eta} +\mathcal{C}_{6}e^{-|k|\eta}\right) \left({\frac{a_{0}}{a}}\right)^{2}\,.  \label{oB1}
\end{equation}
As we will now see, magnetic fields obeying the above evolution law are superadiabatically amplified on the associated scales.

Originally, the amplification effect was discussed on FRW backgrounds containing matter with a specific equation of state (i.e. for~slow-rolling scalar fields and radiation~\cite{TK}). Here we will use a unified approach to show that the effect is essentially independent of the type of matter that fills the universe. To begin with, consider an open FRW cosmology containing a single perfect fluid with barotropic index $w=p/\rho $, where $\rho$ is its energy density and $p$ its isotropic pressure. The scale
factor of such a universe evolves according to the wave-like equation
\begin{equation}
y^{\prime\prime}- \beta^{2}y= 0\,,  \label{y''}
\end{equation}
where $y=a^{\beta}$ and $\beta=(1+3w)/2\neq0$~\cite{B}. The last parameter determines the total gravitational mass of the cosmic medium. When $\beta>0$, in particular, we are dealing with conventional matter. In the opposite case, the fluid has an inflationary equation of state. Solving equation (\ref{y''}) gives
\begin{equation}
y= a^{\beta}= \mathcal{A}_{1}\sinh(\beta\eta)+ \mathcal{A}_{2}\cosh(\beta\eta)\,,  \label{y}
\end{equation}
with $\mathcal{A}_{1,2}$ representing the integration constants. Introducing the ``phase''-parameter $\phi= \tanh^{-1}(\mathcal{A}_{2}/\mathcal{A}_{1})$ and employing some straightforward algebra, solution (\ref{y}) becomes
\begin{equation}
\left({\frac{a}{a_{0}}}\right)^{\beta}= {\frac{\sinh(\beta\eta+\phi)}{\sinh(\beta\eta_{0}+\phi)}}\,, \label{a1}
\end{equation}
where $\beta\eta+\phi>0$~\cite{TCM}. Without loss of generality we may assume that $\phi=0$, which implies that $\beta\eta>0$. Then, we have $\eta>0$ for conventional matter and $\eta<0$ when the medium has negative gravitational mass. In the former case one can easily show that $\lim a_{\eta\rightarrow0^{+}}=0^{+}$ and $\lim a_{\eta\rightarrow+\infty}=+\infty $, at the beginning and at the end of the expansion respectively. When $\beta,\,\eta<0$, on the other hand, we find $\lim a_{\eta\rightarrow-\infty}=0^{+}$ and $\lim a_{\eta\rightarrow0^{-}}=+\infty$. Finally, we may also adopt the normalisation $\sinh(\beta\eta_{0})= 1\Leftrightarrow\eta_{0}=\ln(1+\sqrt{2})/\beta$ and rewrite solution (\ref{a1}) as
\begin{equation}
\left({\frac{a}{a_{0}}}\right)^{\beta}= \sinh(\beta\eta)\,, \label{a2}
\end{equation}
keeping in mind that $\beta\neq0$ and $\beta\eta>0$. Solving the above for the conformal time gives $\eta=\sinh^{-1}[(a/a_{0})^{\beta}]/\beta$, which substituted into the right-hand side of Eq.~(\ref{oB1}) leads to
\begin{eqnarray}
B_{(k)}&=& \mathcal{C}_{5}\left[1+ \sqrt{1+\left({\frac{a_{0}}{a}}\right)^{2\beta}}\;\right]^{|k|/\beta} \left({\frac{a}{a_{0}}}\right)^{|k|-2}\notag\\ &&+\mathcal{C}_{6} \left[1+\sqrt{1+\left({\frac{a_{0}}{a}}\right)^{2\beta}}\;\right]^{-|k|/\beta} \left({\frac{a}{a_{0}}}\right)^{-|k|-2}\,.  \label{oB2}
\end{eqnarray}
This expression monitors the linear evolution of cosmological magnetic fields on spatially open FRW backgrounds that contain a single, poorly conductive perfect fluid with barotropic index $w\neq-1/3$. Solution (\ref{oB2}) shows that $B$-fields coherent on scales close to and beyond the curvature radius of an open Friedmann model (i.e.~with $0<k^{2}<2$) are superadiabatically amplified. When $\beta >0$, for example, Eq.~(\ref{oB2}) reduces to
\begin{equation}
B_{(k)}=\mathcal{C}_{5}\left({\frac{a}{a_{0}}}\right)^{|k|-2}+ \mathcal{C}_{6}\left({\frac{a}{a_{0}}}\right)^{-|k|-2}\,, \label{dust}
\end{equation}
at late times (i.e.~for $a\gg a_{0}$). The same evolution law also holds for non-conventional matter, namely when $\beta<0$, although now at early times (i.e.~for $a\ll a_{0}$). Note that close to the curvature radius (i.e.~as $k^{2}\rightarrow1$) we always find that $B_{(1)}\propto a^{-1}$. Also, the amplification effect grows as we move to increasingly larger wavelengths, with the magnetic decay-rate dropping down to $B_{(1)}\propto a^{\sqrt{2}-2}$ at the homogeneous limit (i.e.~for $k^{2}\rightarrow2$). Inside the curvature radius, on the other hand, the effect of the spatial geometry weakens and the magnetic amplification is less efficient. At the $k^{2}\rightarrow0$ threshold, in particular, the universe is effectively a flat FRW model and we recover the usual adiabatic magnetic decay-rate.

Finally, we should also point out that large-scale magnetic fields are superadiabatically amplified on a Milne background as well. The latter corresponds to an empty, open FRW spacetime with a scale factor that satisfies the condition $a=t$. Expressed in terms of conformal time, this translates to $\mathrm{e}^{\eta}\propto a$. Substituted into solution (\ref{oB1}), the latter ensures decay rates slower than the adiabatic one for all $B$-fields spanning lengths near and beyond the curvature scale of the Milne model. Thus, it appears that the superadiabatic amplification of large-scale magnetic fields is a generic feature of the open FRW universes.

\subsection{The strength of the residual magnetic
field}\label{ssSRMF}
The amount of the magnetic amplification and the strength of the residual field depends on the specifics of the adopted inflationary scenario. Relative to the flat-FRW case, there is also an additional parameter due to the non-Euclidean geometry of the background cosmology. This is the current curvature radius of the universe. Nevertheless, the achieved magnetic growth is quite substantial in almost all the cases of interest. To demonstrate this we essentially repeat the calculation outlined in \S~\ref{ssTRSIMFs}, while taking into account that the magnetic field is now superadiabatically amplified on large scales throughout the evolution of the universe, namely during inflation, reheating, radiation and dust. Staying close to the curvature scale, for simplicity, we have $(\rho_{B}/\rho)_{INF}=f_Q(\rho_{B}/\rho)_{HC}\,\mathrm{e}^{-2N}$ by the end of inflation. Recall that the $HC$-suffix indicates the time of first horizon crossing and $N$ denotes the total number of e-folds between that moment and the end of the inflationary phase. We also remind the reader that $f_Q$ is a dimensionless parameter measuring the efficiency of the quantum mechanical excitation of the electromagnetic modes (with $f_Q\in[0,1]$ -- see also \S~\ref{ssTRSIMFs} earlier). Now, $(\rho_B/\rho)_{RH}\simeq(\rho_B/\rho)_{INF}(M/T_{RH})^{4/3}$ provides the relative change in the energy density of the superadiabatically amplified $B$-field between the end of inflation and that of reheating. Then, we find that
\begin{eqnarray}
r_{RH}= \left({\frac{\rho_{B}}{\rho_{\gamma}}}\right)_{RH}&\simeq& 10^{-54}f_Q\left(\frac{1{\rm Mpc}}{\lambda_{B}}\right)^{2} \left(\frac{M}{10^{14}{\rm GeV}}\right)^4\left(\frac{T_{RH}}{10^{10}{\rm GeV}}\right)^{-2}\,,  \label{r2}
\end{eqnarray}
at the beginning of the radiation era. Comparing this result to Eq.~(\ref{r1}), we can see that the (superadiabatic) magnetic amplification is already substantial by the end of reheating, although it now depends on the energy scale of the adopted inflationary model and on the corresponding reheat temperature. Large-scale $B$-fields, however, are also superadiabatically amplified during the subsequent evolution of the universe. This means that, on lengths close to the curvature scale of the universe, the ratio $r=\rho_{B}/\rho_{\gamma}$ is no longer constant, but increases as $r\propto a^{2}\propto T^{-2}$. Consequently, today we have
\begin{equation}
r_0= r_{RH}\left({\frac{T_{RH}}{T_0}}\right)^{2}\,. \label{r3}
\end{equation}
Our last step is to combine expressions (\ref{r2}) and (\ref{r3}), recalling that $\lambda_{K}= \lambda_{H}/\sqrt{1-\Omega}$ is the curvature scale of a spatially spatially open FRW model. Then, inserting the values $(\rho_{\gamma})_0\sim 10^{-51}~\mathrm{GeV}^{4}$, $T_0\sim10^{-13}$~GeV and $(\lambda_{H})_0\sim10^{3}$~Mpc into the resulting expression, we arrive at
\begin{equation}
\left(\rho_{B}\right)_0\sim 10^{-65}\left(\frac{M}{10^{14}{\rm GeV}}\right)^{4}(1-\Omega_0) \hspace{5mm} \mathrm{GeV}^{4}\,,  \label{B2*}
\end{equation}
having also set $f_Q$ equal to unity. The above immediately translates into
\begin{equation}
B_0\sim 10^{-13}\left(\frac{M}{10^{14}{\rm GeV}}\right)^2 \sqrt{1-\Omega_0} \hspace{5mm} \mathrm{G}\,,  \label{B*}
\end{equation}
which provides the present magnitude of a superadiabatically amplified $B$-field coherent on a scale close to the present curvature radius in an open FRW universe. According to this result, the higher the energy-scale of inflation the stronger the superadiabatic amplification and the residual magnetic field. On the other hand, the closer the current value of the density parameter is to unity (i.e.~the larger the current curvature scale of the universe), the weaker the final $B$-field.

Galactic dynamos require magnetic seeds with strengths between $\sim10^{-12}$ and $\sim10^{-22}$ Gauss when evolved on to present-day values. It has been argued, however, that the lower limit could be pushed down to $\sim10^{-35}$~G in open FRW models~\cite{DLT}. At the same time, current observational data put the value of the $\Omega $-parameter close to unity. In particular, successive reports from WMAP indicate that $|1-\Omega_0|\lesssim10^{-2}$~\cite{Setal}. On these grounds and provided that the universe is open, expression (\ref{B*}) gives
\begin{equation}
B_0\sim 10^{-14}\hspace{5mm}\mathrm{G}  \label{B*1}
\end{equation}
for $M\sim 10^{14}$~GeV and $1-\Omega_0\sim10^{-2}$. Note that the last parameter choice means that the current curvature radius is close to $10^{4}$~Mpc. Clearly, magnetic fields spanning such scales have coherent lengths much larger than 10~Kpc, which is the minimum size required by the dynamo mechanism. Nevertheless, once structure formation starts, these fields should break up and reconnect on scales close to those of  collapsing proto-galactic clouds. We also note that, although $B$-fields close to $10^{-14}$~G can successfully seed the galactic dynamo, they have no observationally significant effect on primordial nucleosynthesis~\cite{CST}, or on the cosmic microwave background spectrum~and isotropy~\cite{BFS}. Finally, we should point out that the presence of magnetic fields with comparable magnitudes (around $10^{-15}$~G) in empty intergalactic space was recently reported in~\cite{TGFBGC}.

The above quoted magnetic strength will increase if we assume higher energy scale for our inflationary model. On the other hand, according to expression (\ref{B*}), the magnetic strength will drop if the current curvature scale is far larger than the Hubble horizon (i.e.~when $1-\Omega_0\ll10^{-2}$). However, the dependence of the residual magnetic strength on $1-\Omega_0$ is fairly weak. This means that magnetic fields capable of supporting the galactic dynamo (i.e.~with $B_0\gtrsim10^{-22}$~G) are possible even when $1-\Omega_0\sim10^{-18}$ (or lower -- when the scale of inflation is higher than $10^{14}$~GeV). Moreover, if the results of~\cite{DLT} are taken at face value, the $1-\Omega_0$ difference can drop down to $\sim10^{-44}$ (or even lower) and still produce $B$-fields able to seed the galactic dynamo (i.e.~with $B_0\gtrsim10^{-35}$~G).\footnote{The dependence of the residual magnetic strength on the current curvature scale of the universe, translates to a dependence on the number of e-folds during the early inflationary phase. In particular, assuming that $\Omega\ll1$ at the beginning of inflation, we have
\begin{equation}
N_{tot}\simeq N_{min}- \ln\left(\sqrt{1-\Omega_0}\,\right)\,,  \label{Ntot}
\end{equation}
where $N_{tot}$ and $N_{min}$ are the total and the minimum number of e-folds respectively~\cite{KT}. Also, typically, around 60 e-folds are necessary for a successful inflationary scenario. Following Eq.~(\ref{Ntot}), open FRW models with $1-\Omega_0\sim10^{-2}$ correspond to $N_{tot}\simeq N_{min}+2$. In this case, the total number of e-folds is marginally larger than the minimum required. Using the same expression, we find that $N_{tot}\simeq N_{min}+20$ for $1-\Omega_0\sim10^{-18}$ and $N_{tot}\simeq N_{min}+50$ when $1-\Omega_0\sim10^{-44}$. Clearly, the larger the number of e-folds, the further away the current curvature scale is pushed and the lower the final magnetic strength.} All these features point towards the same conclusion: that even (very) marginally open FRW universes can sustain cosmological magnetic fields strong enough to support the
galactic dynamo mechanism.

\section{Discussion}\label{sD}
Magnetic fields appear to be ubiquitous in the universe. Despite their widespread presence, however, the origin of the large-scale $B$-fields that we see in galaxies, in galaxy clusters and in remote proto-galactic structures remains an open issue. Given that the galactic dynamo requires an initial magnetic seed in order to operate, the question is where do these seeds come from. The idea of primordial magnetism has certain attractive aspects and gains credence as more evidence for magnetic fields is found at high redshifts. Nevertheless, early-time magnetogenesis faces major problems. The theoretical obstacles are those related to the coherence length-scale and also to the strength of the primordial $B$-field that will seed the galactic dynamo. Inflation can solve the scale-problem, but typically leads to magnetic seeds that are far too weak to be of any astrophysical relevance.

Over the years, several mechanisms producing magnetic fields strong enough to seed the galactic dynamo have appeared in the literature. Almost all of the proposed scenarios, however, operate outside standard cosmology and/or conventional physics. The reason was the adiabatic magnetic decay-law, which was widely believed to hold on all FRW backgrounds, irrespective of the type of matter that they contain and of its electrical properties. Nevertheless, large-scale $B$-fields are also affected by the spatial geometry of the host spacetime. We have shown here that when the FRW models are open, their hyperbolic spatial geometry can superadiabatically amplify magnetic fields that are coherent on scales close to and beyond the curvature radius of such a universe. We have also shown that the amplification effect is essentially independent of the material content of the universe and appears to be a generic feature of the open FRW spacetimes. Moreover, the strength of the residual $B$-field is substantially stronger than the one typically associated with a flat Friedmann model, as it varies between $10^{-20}$ and $10^{-10}$ Gauss for a fairly broad range of initial conditions. In practise, all these mean that even a marginally open FRW universe (with $1-\Omega_0\sim10^{-2}-10^{-18}$ today -- and potentially much less) can produce and sustain inflationary magnetic fields capable of successfully seeding the galactic dynamo.

\end{document}